\documentclass[twocolumn,prb,showpacs]{revtex4}

\usepackage{graphicx}
\usepackage{dcolumn}
\usepackage{bm}

\begin{document}

\title{Biordered superconductivity and strong pseudogap state}

\author{V.~I.~Belyavsky and Yu.~V.~Kopaev}

\address {P.~N.~Lebedev Physical Institute of Russian Academy of
Sciences, Moscow, 119991, Russia}

\begin{abstract}
Interrelation between the two-particle and mean-field problems is
used to describe the strong pseudogap and superconducting states
in cuprates. We present strong pseudogap state as off-diagonal
short-range order (ODSRO) originating from quasi-stationary states
of the pair of repulsing particles with large total momentum
($K$~-~pair). Phase transition from the ODSRO state into the
off-diagonal long-range ordered (ODLRO) superconducting state is
associated with Bose-Einstein condensation of the $K$~-~pairs. A
checkerboard spatial order observable in the superconducting state
in the cuprates is explained by a rise of the $K$~-~pair density
wave. A competition between the ODSRO and ODLRO states leads to
the phase diagram typical of the cuprates. Biordered
superconducting state of coexisting condensates of Cooper pairs
with zero momentum and $K$~-~pairs explains some properties of the
cuprates observed below $T^{}_c$: Drude optical conductivity,
unconventional isotope effect and two-gap quasiparticle spectrum
with essentially different energy scales.

\end{abstract}


\maketitle

\section{Introduction}

The high-$T^{}_c$ cuprate superconductor can be considered a doped
two-dimensional (2D) antiferromagnetic (AF) insulator with strong
Coulomb repulsion. \cite{Anderson} Most commonly, one accepts that
the physics of the cuprates can be understood within the simplest
one-band models of strong correlated systems such as the Hubbard
and $t-J$ models. It seems impossible to obtain analytic solutions
for the ground states of these 2D models. Therefore, a variational
approach based on a choice of an appropriate trial wave function
can be considered as a natural way to solve the cuprate problem.
\cite{six_band}

The wave function of the resonating valence bond (RVB) ground
state \cite{Anderson} derived from Gutzwiller projected
\cite{Gutzwiller} $d$~-~wave ground state of the
Bardeen-Cooper-Schrieffer (BCS) model eliminates a possibility of
double occupancy of a site ({\it{no double-occupancy constraint}})
and corresponds to extremely strong on-site correlations. A great
many of important results, obtained within the RVB approach, can
be considered as a ground of the physics of the cuprates.

However, the problem is complicated by the fact that, in a strong
correlated system, ground state energies of different ordered
states may be close to each other: \cite{Sachdev} $d$~-~wave
superconductor (dSC), staggered flux phase, \cite{Affleck} spin
and charge density waves (SDW and CDW, respectively) and some
others. \cite{Nayak} Account of the competition and coexistence of
such ordered states within extremely simplified models by
numerical tools leads to a wide variety of phase diagrams. It is
clear that, among them, one can always find the diagrams reminding
those typical of the cuprates.

Various approximations within the on-site Coulomb repulsion models
often lead to antipodal conclusions concerning a possibility of
the superconducting (SC) state itself. \cite{Laughlin} The
complete suppression of the double occupancy under Gutzwiller's
projection promotes a rise of an insulating rather than SC order.
In this connection, Laughlin \cite{Laughlin} has proposed an
alternative approach to clarify the problem of the
superconductivity of the cuprates. Instead of numerical study of a
highly simplified Hamiltonian, he suggested to select a reasonable
ground state in order to determine the Hamiltonian leading to such
a state.

To take into account a realistic on-site repulsion, the ground
state of Laughlin's gossamer superconductor \cite{Laughlin} is
chosen in the form of an incomplete projected BCS $d$~-~wave state
with partially suppressed double occupancy. Hamiltonian with such
an exact forethought ground state, along with strong on-site
repulsion, manifests an attractive term which can lead to dSC
state. \cite{Laughlin}

Gossamer superconductivity of the underdoped compound can be
associated with a band of states with relatively low spectral
weight inside a pronounced insulating forbidden band so that the
chemical potential turns out to be pinned near the middle of this
band. \cite{Laughlin}

The repulsion-induced dSC state is highly sensitive to electron
dispersion. \cite{LNW} The simplest tight-binding approximation
taking into account only the nearest neighbors ($t$~-~model) seems
to be insufficient, therefore, it is necessary to consider more
complicated dispersion with the next nearest neighbor terms
($t-t^{\prime}$~-~model). Numerical study shows \cite{LNW} that
the stable dSC state corresponds to hopping integral ratio
$t_{}^{\prime}/t$ within a narrow range near $t_{}^{\prime}/t
=-0.3$. Just the same value of $t_{}^{\prime}/t$ is consistent
with available angle resolved photoemission spectroscopy (ARPES)
data. \cite{Damascelli}

As follows from the $SU(2)$ approach to the RVB problem,
\cite{LNW} it is necessary to consider doublets of fermions and
bosons to realize the spin-charge separation correctly. \cite{ILW}
Two minima of the $SU(2)$ boson dispersion are relative to the
points ($0,0$) and (${\pi},{\pi}$) in the 2D Brillouin zone,
\cite{ILW} therefore, the SC pairing channel corresponding to
large pair momentum should be taken into account along with the
Cooper channel corresponding to zero pair momentum.

Geshkenbein et al. \cite{GIL} have assumed that an enhancement of
the scattering between the saddle points of electron dispersion
results in the fact that the electron-electron interaction with
large momentum transfer can be ``less repulsive'' with respect to
small transfer. Therefore, in the vicinities of the saddle points,
fermions may be paired into bosons. Such noncoherent preformed
pairs arising near the antinodal arcs of the Fermi contour (FC)
might exist in the pseudogap state of underdoped cuprates as a
normal Bose liquid. \cite{GIL}

The FC outside of the arcs corresponds to unpaired fermions
coexisting with the preformed pairs. Bose-Einstein condensation
(BEC) of the preformed pairs with large momentum due to their
interaction with unpaired particles results in the SC gap on the
whole of the FC. The SC state that arises in such a way describes
reasonably rather wide (intermediate with respect to BCS and BEC
limiting cases) fluctuation region above $T^{}_c$.

Instabilities in 2D strong correlated electron system were
investigated within the $t-t^{\prime}$~--~model at small
$t_{}^{\prime}$ by renormalization-group (RG) methods
\cite{Furukawa,Honerkamp} using a discretization of the FC into a
finite number of patches. The singularity in the Cooper channel
exhibits a squared logarithmic divergence at low energies. For
insulating Peierls channel with electron-hole pair momentum
${\bm{Q}}^{}_{\pi}={({\pi},{\pi})}$, the singularity also exhibits
a squared logarithm in the particular case $t_{}^{\prime}=0$ when
nested FC has the form of a square coinciding with the boundary of
the magnetic Brillouin zone of the parent compound. At
$t_{}^{\prime}\neq 0$ and low doping, that is in the case of a
deviation of the FC from the perfect nesting, the divergence is
found more weak with logarithmic enhancement of the order of
${\ln{|t/t_{}^{\prime}}|}$ under the condition that
$|t_{}^{\prime}|\ll |t|$. The singularities in the insulating and
SC channels corresponding to zero and ${\bm{Q}}^{}_{\pi}$ pair
momenta, respectively, are found to be logarithmic in the case of
small but nonzero $|t_{}^{\prime}/t|$.  Such a case corresponds to
approximately nested FC disposed close to saddle-point van Hove
singularity. RG approach, involving the nesting effects,
\cite{ZYD} gives a possibility to select singular contributions
into pairing channels but remains corresponding pre-exponential
factors to be undetermined.

General symmetry consideration, based on Zhang's SO(5) theory
\cite{Zhang} or SU(4) theory by Guidry et al., \cite{Guidry} shows
that one should take into account a closed set of competing
ordered states to describe key features of correlated electron
system. In this sense, singlet SC pairing channel with large
momentum incorporating singlet orbital insulating long-range
(possibly, hidden) \cite{Varma,CLMN} or short-range (fluctuating
between dSC and staggered flux states) \cite{Lee} order may be
naturally connected with the Cooper channel. Thus, there should be
two SC gap parameters related to large and zero pair momentum,
respectively.

The SC gap, which determines $T^{}_c$ and corresponds to a rise of
the coherence in the system of electron pairs, can be directly
extracted from experiments on Andreev reflection \cite{Deutscher}
or Josephson tunneling. \cite{Joseph} The observation of two SC
gaps of about $10$~$meV$ and $50$~$meV$, respectively, in tunnel
experiment \cite{Ved} in Bi2212 (in particular, a suppression of
the lesser gap in high magnetic field at temperatures
$30-50$~$mK$) may be considered as an indirect evidence in favour
of two SC energy scales in the cuprates.

One more energy scale, observed in ARPES and tunnel spectra of
underdoped cuprates, \cite{Damascelli,Timusk} can be associated
with the strong pseudogap state. \cite{Norman} To describe this
state one can start from a reasonably chosen one-particle Green
function. Recently, Yang et al. \cite{YRZ} developed RVB
phenomenology of the pseudogap state based on the assumption that
this state can be viewed as a liquid formed by an array of weakly
coupled two-leg Hubbard ladders. The coherent part of the Green
function obtained within the random phase approximation is
consistent with the Luttinger theorem \cite{Konik} and describes
evolution of the FC (from small pockets to closed contour) with
doping. Similar results follows from both the spin-charge
separation approach \cite{Ng} and phenomenological account of
short range insulating order above $T^{}_c$. \cite{Sad}

In this paper, we develop the concept of Coulomb pairing
\cite{BK_UFN} leading to the {\it{biordered}} state originating
from two SC pairing channels, with large and zero pair momenta.
The all-sufficient conditions of repulsion-induced
superconductivity in these two channels are discussed in Sec.~II.
Sec.~III deals with the strong pseudogap state arising from
incoherent quasi-stationary states of pairs with large momenta
that are inherent in the screened Coulomb pairing potential. In
Sec.~IV, we consider the symmetry and two-gap spectrum of the
biordered state. Finally, some possible manifestations of the
strong pseudogap and biordered SC states are discussed in Sec.~V.

\section{Competing pairing channels}

Screening of Coulomb repulsion in three-dimensional isotropic
degenerate electron gas results in momentum dependent interaction
energy of two electrons,
\begin{equation}\label{1}
U(k)=4{\pi}e_{}^2/[k_{}^2\epsilon (k)],
\end{equation}
where static permittivity has the form \cite{Ziman}
\begin{equation}\label{2}
\epsilon (k)=1+{\frac{k_0^2}{2k_{}^2}} \left
(1+{\frac{1-x_{}^2}{4x}}\ln{\left |{\frac{1+x}{1-x}}\right
|}\right ).
\end{equation}
Here, $x=k/2k^{}_F$, $k^{}_F$ and $k^{-1}_0=(4{\pi}e^2_{}n
g)_{}^{1/2}$ are Fermi momentum and screening length,
respectively, $n$ and $g$ are electron concentration and density
of states on the Fermi level.

Kohn singularity at $k=2k^{}_F$ leads to the interaction energy
with damped Friedel oscillation in the real space. At a distance
$r\gg k_F^{-1}$, this potential can be written as \cite{Ziman}
\begin{equation}\label{3}
U(r)\simeq {\frac{e^2_{}}{2{\pi}}}{\frac{\cos{2k^{}_Fr}}{r^3_{}}}.
\end{equation}
Kohn and Luttinger \cite{KL} have argued that attractive
contribution into screened Coulomb repulsion originating from
Friedel oscillation is sufficient to ensure Cooper pairing with
non-zero angular momentum. Because of the weakness of the Kohn
singularity, corresponding SC transition temperature turns out to
be very low. \cite{KL}

In the case of nested FC, the Kohn singularity transforms into the
Peierls one with strong anisotropy of $\epsilon ({\bm{k}})$.
Therefore, effective pairing interaction can be enhanced both in
particle-hole and particle-particle channels. In particular, this
can give rise to CDW or SDW in singlet or triplet insulating
pairing channels, respectively.

Peierls singularity in a particle-hole channel originates from the
fact that momentum transfer turns out to be equal to nesting
vector ${\bm{Q}}$ for any particle on the FC (Fig.~1a). For the
sake of simplicity, in Fig.~1 the FC is presented as a square
corresponding to the $t$~-~model at half-filling. In the case of
low doping and under the condition that $|t_{}^{\prime}|\ll t$,
one can expect that the FC may be found close to this square.
\cite{Furukawa}

In the Cooper channel, Peierls enhancement of the Kohn singularity
emerges as appreciably more weak because momenta before and after
scattering (${\bm{p}}$ and ${\bm{p}}_{}^{\prime}$ in Fig.1b,
respectively) giving rise to this enhancement should be related as
${\bm{p}}-{\bm{p}}_{}^{\prime}\approx{\bm{Q}}$. Integration over
momentum transfer ${\bm{p}}_{}^{\prime}-{\bm{p}}$ along the whole
of nested FC smoothes down the Peierls singularity from the Cooper
channel. Logarithmic singularity
(${\ln{(|\varepsilon}^{}_0/{\varepsilon}|)}$, where
${\varepsilon}^{}_0$ is a cut-off energy) in the Cooper channel is
ensured by a general feature of electron dispersion,
${\varepsilon}(-{\bm{p}})={\varepsilon}({\bm{p}})$, that holds for
any momentum ${\bm{p}}$ (statistical weight of the Cooper pair is
proportional to the length of the entire FC).

Density of states of 2D system manifests logarithmic van Hove
singularities originating from saddle-point vicinities with
hyperbolic metric. Due to close proximity of the FC and the
isoline connecting saddle points ($\pm{\pi},\pm{\pi}$), effective
coupling constant $w$ turns out to be logarithmically enhanced,
$w\rightarrow w\cdot
{\ln{(2tk^2_c/|{\varepsilon}-{\varepsilon}^{}_s|)}}$.
\cite{Furukawa} Here, ${\varepsilon}^{}_s$ is saddle point energy
(within the $t$~-~model, ${\varepsilon}^{}_s=0$) and $k^{}_c$ has
meaning of a scale of the part of 2D Brillouin zone with
hyperbolic metric.

\begin{figure}
\includegraphics[]{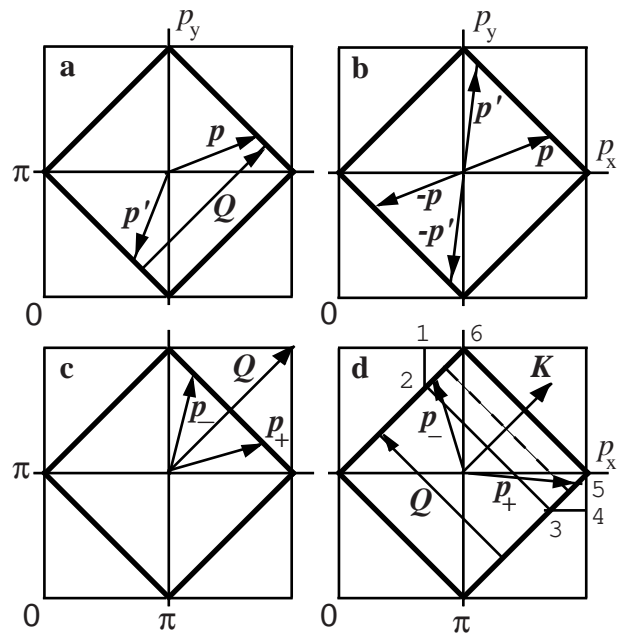}
\caption[*]{Nested Fermi contour (bold line) in the form of a
square coinciding with 2D magnetic Brillouin zone of parent
compound with halh-filled conduction band (within tight-binding
model with nearest-neighbor interactions). {\bf{a}}: electron-hole
pairing with momenta ${\bm{p}}$ and
${\bm{p}}_{}^{\prime}={\bm{p}}-{\bm{Q}}$ (${\bm{Q}}$ is nesting
momentum); {\bf{b}}: Cooper pairing with zero total momentum
(${\bm{p}}$ and ${\bm{p}}_{}^{\prime}$ are momenta before and
after scattering, respectively); {\bf{c}}: SC pairing with nesting
momentum (${\bm{p}}^{}_{\pm}={\bm{Q}}/2\pm {\bm{k}}$, where
${\bm{k}}$ is momentum of the relative motion of the pair). Domain
of kinematic constraint is degenerated into a line coinciding with
one of the sides of the square; {\bf{d}}: SC pairing with an
incommensurate total momentum ${\bm{K}}$
(${\bm{p}}^{}_{\pm}={\bm{K}}/2\pm {\bm{k}}$). Domain of kinematic
constraint is bounded by the line 1-2-3-4-5-6. Parts 2-6 and 3-5
of this line are nested pieces of the FC resulting in singular
contribution into this pairing channel.}\label{F1.eps}
\end{figure}

In the case of SC pairing with large total momentum ${\bm{K}}$
(${\bm{K}}$~-~channel), the momenta of the particles composing a
pair with given momentum (${\bm{K}}$~-~pair), both being either
inside or outside the FC at $T=0$, should belong to only a part of
the Brillouin zone (domain of {\it{kinematic constraint}}) rather
than the whole one. In a general case, kinetic energy of the pair
with relative motion momentum ${\bm{k}}$,
\begin{equation}\label{4}
2{\xi}^{}_{{\bm{K}}}({\bm{k}})={\varepsilon}({\bm{K}}/2+{\bm{k}})+
{\varepsilon}({\bm{K}}/2-{\bm{k}})-2{\mu},
\end{equation}
vanishes only at some points of the FC inside this domain ($\mu$
is the chemical potential). Therefore, in contrast with the Cooper
pairing when ${\xi}^{}_0({\bm{k}})=0$ on the whole of the FC,
integration over ${\bm{k}}$ eliminates the logarithmic singularity
in the ${\bm{K}}$~-~channel. However, if kinetic energies,
${\varepsilon}({\bm{K}}/2+{\bm{k}})$ and
${\varepsilon}({\bm{K}}/2-{\bm{k}})$ coincide on finite pieces of
the FC (``pair'' Fermi contour, PFC), logarithmic singularity
${\ln{(|\varepsilon}^{}_0/|{\varepsilon}|)}$ survives and the
${\bm{K}}$~-~channel can result in the SC order. \cite{BK_UFN}
{\it{Mirror nesting}} condition,
\begin{equation}\label{5}
{\varepsilon}({\bm{K}}/2+{\bm{k}})
={\varepsilon}({\bm{K}}/2-{\bm{k}}),
\end{equation}
determines the locus in the momentum space that logarithmically
contributes to the ${\bm{K}}$~-~channel. Statistical weight of
${\bm{K}}$~-~pair, proportional to the length of the PFC, can be
less in comparison with the Cooper channel.

Mirror nesting is a necessary (not all-sufficient) condition of
the SC pairing with large momentum. Indeed, this condition is
perfectly satisfied in the case of nested FC when
${\bm{K}}={\bm{Q}}^{}_{\pi}$ (Fig.~1c). However, it is obvious
that, in such a case, domain of kinematic constraint degenerates
into a line resulting in zero statistical weight of the paired
state. To obtain finite statistical weight, one can choose
incommensurate pair momentum ${\bm{K}}\neq {\bm{Q}}^{}_{\pi}$.
This results in the domain of kinematic constraint in the form of
relatively narrow strip containing the PFC as shown in Fig.~1d.
The coefficient of the logarithmic contribution to the
${\bm{K}}$~-~channel should be proportional to the length of the
PFC. Considering the PFC as two patches connected by nesting
vector ${\bm{Q}}^{}_{\pi}$ (Fig.~1d), one can conclude that this
coefficient has to be logarithmically enhanced by umklapp
scattering inherent in the Peierls channel. \cite{Furukawa} Patch
approximation \cite{Furukawa} appears to be relatively good just
in the case of short PFC. The reason is that integration over
momentum transfer ${\bm{p}}-{\bm{p}}_{}^{\prime}\approx
{\bm{Q}}^{}_{\pi}$ cannot completely eliminate enhancement of the
pairing interaction due to logarithmic singularities of the
permittivity.

Another necessary condition of SC pairing under repulsion is
connected with the existence of oscillating attractive
contribution into the pairing potential. It should be noted that
an oscillation itself cannot ensure a rise of a bound state. For
example, simple step-wise repulsive potential
$U({\bm{k}})=U^{}_0>0$ defined in a finite domain of the momentum
space oscillates in the real space. However, by analogy with the
problem of a bound state in one-dimensional asymmetric potential
well, \cite{LL_III} such a potential cannot result in a bound
state even under mirror nesting.

One can consider screened Coulomb potential
$U({\bm{k}}-{\bm{k}}_{}^{\prime})$ as a kernel of Hermite integral
operator with complete orthonormal system of eigenfunctions
defined within domain of kinematic constraint ${\Xi}$,
\begin{equation}\label{6}
{\varphi}^{}_s({\bm{k}})={\lambda}^{}_s\sum_{{\bm{k}}_{}^{\prime}\in{\Xi}}
U({\bm{k}}-{\bm{k}}_{}^{\prime})
{\varphi}^{}_s({\bm{k}}_{}^{\prime}).
\end{equation}
Here, a set of ${\lambda}^{}_s$ represents the spectrum of a
pairing operator which can be written in the form of the
Hilbert-Schmidt expansion,
\begin{equation}\label{7}
w({\bm{k}},{\bm{k}}_{}^{\prime})=\sum_s{\frac{{\varphi}^{}_s({\bm{k}})
{\varphi}^{\ast}_s({\bm{k}}_{}^{\prime})}{{\lambda}^{}_s}}.
\end{equation}
The necessary (and sufficient, under mirror nesting) condition of
the SC pairing under repulsion is the existence of at least one
{\it{negative eigenvalue}} of the pairing operator. \cite{BK_UFN}

In the case of comparatively small domain of kinematic constraint,
one can replace the screened Coulomb potential by its expansion in
powers of momentum transfer,
${\bm{\kappa}}={\bm{k}}-{\bm{k}}_{}^{\prime}$, up to the term of
the second order,
\begin{equation}\label{8}
w({\kappa})=U^{}_0r^2_0(1-{\kappa}_{}^2r^2_0/2)
\end{equation}
where $U^{}_0$ and $r^{}_0$ have meaning of an on-site repulsive
energy and screening length, respectively. The simplest repulsive
kernel (\ref{8}), defined inside ${\Xi}$, has two even and two odd
(with respect to inversion ${\bm{k}}\rightarrow -{\bm{k}}$)
eigenfunctions. Singlet SC order parameter should be determined by
only even eigenfunctions belonging to eigenvalues ${\lambda}^{}_1$
and ${\lambda}^{}_2$ of opposite sign.

Scattering between nested pieces of the FC leads to strong
anisotropy of the permittivity. Therefore, expansion of
$w({\bm{k}},{\bm{k}}_{}^{\prime})$ in powers of momentum transfer
close to nesting momentum ${\bm{Q}}$ appears to be anisotropic as
well. Resulting pairing interaction kernel, analogous to
Eq.(\ref{8}), preserves its eigenvalue feature
${\lambda}^{}_1{\lambda}^{}_2<0$. It should be noted that nesting
momentum ${\bm{Q}}\neq {\bm{Q}}^{}_{\pi}$, therefore, ${\bm{Q}}$
has the meaning of a new nesting momentum, inherent in the real
FC, which can result in the Peierls enhancement of the SC pairing.

Matrix of pairing operator Eq.~(\ref{7}) between its
eigenfunctions is diagonal, $w^{}_{ss_{}^{\prime}}
={\lambda}^{-1}_s {\delta}^{}_{ss_{}^{\prime}}$. The necessary
condition of the existence of nontrivial solution to the
self-consistency equation with kernel (\ref{8}) has the form
${\lambda}^{}_1{\lambda}^{}_2<0$. Written in arbitrary basis, it
takes the form of the Suhl inequality,
\begin{equation}\label{9}
w^{}_{11}w^{}_{22}-w^{}_{12}w^{}_{21}<0,
\end{equation}
introduced as a necessary condition of superconductivity within
two-band model. \cite{Suhl}

The effective pairing potential oscillates in the real space and,
in agreement with Laughlin's proposal, \cite{Laughlin} manifests
repulsive core at small distance (corresponding to incomplete no
double-occupancy constraint) and attractive contribution outside
of the core. Thus, there is a possibility of a rise not only of a
bound (with negative energy, $E<0$) but also of quasi-stationary
(with $E>0$) paired state with large momentum (Fig.~2).

The singular contribution into SC order parameter is determined by
relatively small vicinity of the PFC with an energy scale
${\varepsilon}^{}_0$. In this respect, repulsion-induced
${\bm{K}}$~-~pairing seems to be similar to phonon-mediated
pairing arising from attraction with negative coupling constant
$V$. Rough estimation \cite{BTSh} of Coulomb repulsion within the
phonon-mediated mechanism of superconductivity leads to the fact
that, to ensure SC pairing, $|V|$ must exceed a threshold value,
\begin{equation}\label{10}
|V|>{\frac{U^{}_c}{1+gU^{}_c\ln{({E^{}_F}/{\varepsilon}^{}_D)}}}.
\end{equation}
Here, $E^{}_F$ and $U^{}_c$ are Fermi and average Coulomb
energies, respectively. Phonon-mediated attraction is defined
inside a narrow layer (domain of dynamic constraint with energy
scale of the order of Debye energy ${\varepsilon}^{}_D$)
enveloping the FC.

\begin{figure}
\includegraphics[]{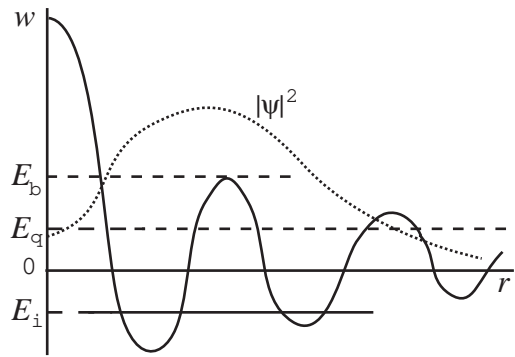}
\caption[*]{Repulsive pairing potential, $w(r)$, and bound state,
$|{\psi}|_{}^2$ (dotted line), distributions in the real-space
(schematically). Energies $E^{}_i$ and $E^{}_q$ correspond to
bound and quasi-stationary states, respectively. Barrier height
$E^{}_b$ corresponds to a break of the pair without tunnelling
through the barrier.}\label{F2.eps}
\end{figure}

One can expect that a deviation from perfect mirror nesting
condition (\ref{5}) outside of the PFC does not eliminate the
logarithmic singularity in the ${\bm{K}}$~-~channel. Typical of
high~-~$T^{}_c$ cuprates, nearly nested FC in the form of a square
with rounded corners \cite{Damascelli} and finite screening length
of the Coulomb interaction result in a weak suppression of the
singularity. The SC gap turns out to be preserved on the PFC (by
analogy with the case of insulating pairing studied by Losovik and
Yudson). \cite{LYu}

The Cooper and ${\bm{K}}$~-~channels can be found as approximately
equally enhanced by both Peierls singularity of screening and
proximity of the FC to van Hove singularity of the density of
states. Therefore, it is the kinematic constraint that can ensure
the preference of the ${\bm{K}}$~-~channel over the Cooper
channel. One can believe that, in doped cuprates, the singular
contribution to Peierls enhanced Coulomb ${\bm{K}}$~-~pairing can
dominate the Cooper channel.

\section{Strong pseudogap state}

Taking into account the ground state instability due to a rise of
pairs, the mean-field approach to the problem of superconductivity
excludes fluctuations of paired states from consideration. Within
the mean-field theory, \cite{BCS} the SC gap is directly relative
to the binding energy of a pair resulted from the two-particle
Cooper problem. \cite{Cooper} In the case of the
${\bm{K}}$~-~pairing, such a problem may admit more complicated
solution as compared to the attraction-induced pairing.
\cite{BK_UFN}

Integral equation which determines a wave function of the relative
motion of two interacting particles (holes) above (below) the FC
can be written as
\begin{equation}\label{11}
{\psi}({\bm{k}})=G^{(0)}_{}({\omega};{\bm{k}})\sum_{{\bm{k}}_{}^{\prime}\in{\Xi}}
w({{\bm{k}},{\bm{k}}_{}^{\prime}}){\psi}({\bm{k}}_{}^{\prime}).
\end{equation}
Here, ${\omega}$ and ${\bm{k}}$ (${\bm{k}}_{}^{\prime}$) are an
energy and momentum of the relative motion before (after)
scattering, respectively, and
\begin{equation}\label{12}
G^{(0)}_{}({\omega};{\bm{k}})=[{\omega}-{\xi}({\bm{k}})+i\,{\gamma}]_{}^{-1}
\end{equation}
is one-particle Green function corresponding to free relative
motion of the ${\bm{K}}$~-~pair, ${\gamma}\rightarrow +0$. In
contrast with one-particle Landau Fermi liquid Green function, the
condition that $[G_{}^{(0)}(0,{\bm{k}})]_{}^{-1}=0$ does not
determine a closed FC. Indeed, a locus in the momentum space
resulting from this condition written in equivalent form
${\xi}^{}_{\bm{K}}({\bm{k}})=0$, is either some isolated points or
finite pieces of mirror nested FC.

In the case of mirror nested FC, one can separate a singular
contribution to the Green function originating from relatively
small (with energy scale ${\varepsilon}^{}_0$) part ${\Xi}^{}_s$
of the domain of kinematic constraint ${\Xi}$. The rest of the
domain, including an energy range from ${\varepsilon}^{}_0$ up to
a cut-off value of about $E^{}_F$, results in a regular
contribution into $G^{(0)}_{}$. One can consider this contribution
in a way similar to the account for the Coulomb repulsion within
phonon-mediated pairing attraction scenario \cite{BTSh} and
renormalize kernel Eq.~(\ref{7}) to a kernel, defined inside
${\Xi}^{}_s$, with the same spectrum.

One-particle Green function $G({\omega};{\bm{k}})$ corresponding
to relative motion of ${\bm{K}}$~-~pair of particles (holes)
excited above (below) the FC can be represented in the basis
formed by the eigenfunctions of the renormalized pairing operator
$w({\bm{k}},{\bm{k}}_{}^{\prime})$,
\begin{equation}\label{13}
G^{}_{ss_{}^{\prime}}({\omega})=\sum_{{\bm{k}}\in
{\Xi}}{\varphi}_s^{\ast}({\bm{k}})G({\omega};{\bm{k}})
{\varphi}_{s_{}^{\prime}}^{}({\bm{k}}).
\end{equation}
Matrix elements (\ref{13}) are the solutions to Dyson equation,
\begin{equation}\label{14}
\sum_{s_{}^{{\prime}{\prime}}}\{{\delta}^{}_{ss_{}^{{\prime}{\prime}}}
-{\lambda}^{-1}_{s_{}^{{\prime}{\prime}}}G^{(0)}_{ss_{}^{{\prime}{\prime}}}
({\omega})\}G^{}_{s_{}^{{\prime}{\prime}}s_{}^{\prime}}({\omega})=
G^{(0)}_{ss_{}^{{\prime}}}({\omega}),
\end{equation}
in which matrix elements $G^{(0)}_{ss_{}^{{\prime}}}({\omega})$ of
free Green function (\ref{12}) are defined similar to
Eq.(\ref{13}).

Pairing operator with two even eigenfunctions \cite{BKTS} results
in $2\times 2$ matrix (\ref{13}). One can resolve  Eq.~(\ref{14})
with respect to $G^{}_{ss_{}^{{\prime}}}({\omega})$ and then
obtain $G({\omega};{\bm{k}})$ in the form
\begin{equation}\label{15}
G({\omega};{\bm{k}})=D_{}^{-1}({\omega})[G^{(0)}_{}
({\omega};{\bm{k}})-B({\omega};{\bm{k}})],
\end{equation}
where
\begin{equation}\label{16}
B({\omega};{\bm{k}})=
{\lambda}^{-1}_1{\lambda}^{-1}_2B({\omega})\sum_{s=1}^2{\lambda}^{}_s
|{\varphi}_s^{}({\bm{k}})|_{}^2 ,
\end{equation}
\begin{equation}\label{17}
B({\omega})=G^{(0)}_{11}({\omega})G^{(0)}_{22}({\omega})-
G^{(0)}_{12}({\omega})G^{(0)}_{21}({\omega}),
\end{equation}
and
\begin{equation}\label{18}
D({\omega})=1-\frac{G^{(0)}_{11}({\omega})}{{\lambda}^{}_1}-
\frac{G^{(0)}_{22}({\omega})}{{\lambda}^{}_2}+ \frac{
B({\omega})}{{\lambda}^{}_1{\lambda}^{}_2} .
\end{equation}

In the case of mirror nested FC, Green function (\ref{15})
manifests a pole resulting in a bound state with negative energy
${\omega}=E^{}_i$ determined from equation $D({\omega})=0$ in
which all functions $G^{(0)}_{ss_{}^{\prime}}({\omega})$ are real.
This pole is related to instability of the ground state with
respect to a rise of pairs. Within a small vicinity of the pole,
Green function (\ref{15}) can be represented as
\begin{equation}\label{18'}
G({\omega};{\bm{k}})={\frac{[G^{(0)}_{}
({\omega};{\bm{k}})-B({\omega};{\bm{k}})]}{D_{}^{\prime}(E_i^{})}}
{\frac{1}{{\omega}-E^{}_i}}
\end{equation}
where $D_{}^{\prime}=d\;D/d\;{\omega}$.

At ${\omega}>0$, Green functions (\ref{13}) are complex and
equation $D({\omega})=0$ can lead to complex solution
${\omega}=E^{}_q-i{\Gamma}$ where $E^{}_q$ and ${\Gamma}$ have
meanings of energy and decay of quasi-stationary state (QSS) of
the relative motion of ${\bm{K}}$~-~pair, respectively.
\cite{BKTS} Near this complex pole, Green function (\ref{15}) has
the form of Eq.~(\ref{18'}) where $E^{}_i$ should be replaced by
$E^{}_q-i{\Gamma}$.

Wave functions of the relative motion of ${\bm{K}}$~-~pair
corresponding to both bound state and QSS, are localized, in main,
in a wide region of the real space outside the repulsive core as
shown schematically in Fig.~2.

The ${\bm{K}}$~-~pairs can exist above $T^{}_c$ as long-living QSS
due to considerable increase of density of states in a narrow
vicinity of $E^{}_q$. To overcome the potential barrier before
tunnel decay, such a non-coherent pair should accumulate an energy
exceeding barrier height $E^{}_b$. Thus, the energy
$E^{}_q-E^{}_i$ is sufficient to destroy SC coherence whereas
corresponding pair-break energy should exceed $E^{}_b-E^{}_i$. A
temperature range between the SC transition temperature
$T^{}_c\sim E^{}_q-E^{}_i$ and a crossover one,
$T^{\ast}_{str}\sim E^{}_b-E^{}_i$, can be interpreted as a strong
pseudogap state observable above $T^{}_c$ in underdoped cuprates.
If density-of-states peak at ${\omega}\approx E^{}_q$ turns out to
be smoothed due to ${\Gamma}$ being large enough, the strong
pseudogap state becomes unobservable. In such a case, the SC
transition from coherent into non-coherent state should be
assisted with a break of pairs at energies $\approx E^{}_b-E^{}_i$
similar to that in the BCS theory.

By analogy with the interrelation between the Cooper two-particle
problem \cite{Cooper} and BCS theory, \cite{BCS} the pair-break
energy $E^{}_b-E^{}_i$ due to direct excitation of particles from
bound state into continuous spectrum should be transformed into
momentum-dependent energy gap ${\Delta}({\bm{k}})$ in the
quasiparticle spectrum. In the strong pseudogap state, this gap,
due to a non-coherence of QSS, can be presented as
${\Delta}={\sqrt{{\Delta}^2_{c} +{\Delta}^2_{p}}}$. Here,
${\Delta}^{}_{c}\sim E^{}_q-E^{}_i$ corresponds to transition from
the coherent into non-coherent QSS and ${\Delta}^{}_{p}\sim
E^{}_b-E^{}_q$ can be related to a break of ${\bm{K}}$~-~pair as a
result of transition between two non-coherent states.

Microscopically, SC gap ${\Delta}^{}_{c}$ and strong pseudogap
${\Delta}^{}_{p}$ emerge with random phases. Therefore, mean-field
value ${\Delta}^{}_{p}$ vanishes at any temperature whereas
${\Delta}^{}_{c}$ becomes nonzero below $T^{}_c$ due to Bose
condensation of ${\bm{K}}$~-~pairs from QSS into the bound state
and vanishes only above $T^{}_c$. However, nonzero mean square
strong pseudogap, $|{\Delta}^{}_{p}|_{}^2\neq 0$, may become
apparent well above $T^{}_c$. In this sense, pseudogap parameter
${\Delta}^{}_{p}$, corresponding to decay of QSS of
${\bm{K}}$~-~pairs, reminds RVB spin liquid pseudogap introduced
by Yang et al. \cite{YRZ} However, it has different physical
meaning.

Green function $G^{(0)}_{}(0;{\bm{k}})$ changes sign on the PFC
from positive to negative through an infinity. Therefore, Green
function (\ref{15}) at ${\omega}=0$ manifests the same feature. It
should be noted that, in the case of Cooper pairing, the Green
function changes sign on the whole of the FC. \cite{YRZ} In
addition, Green function $G^{(0)}_{}(0;{\bm{k}})$ changes sign on
a zero line determined by equation $G^{(0)}_{}(0;{\bm{k}})
=B(0;{\bm{k}})$. This line does not coincide with the PFC.

Green function (\ref{15}) of the two-particle problem has a pole
corresponding to a bound state of the relative motion of the pair.
Therefore, one can suppose, in line with Yang et al., \cite{YRZ} a
phenomenological BCS-like form of the coherent contribution to the
normal (diagonal) Gor'kov Green function of the mean-field problem
\cite{AGD}
\begin{equation}\label{19}
G^{}_{}({\omega};{\bm{k}})=z^{}_{\bm{k}} \left
[{\frac{u^2_+({\bm{k}})} {{\omega}-E({\bm{k}})+i{\Gamma}}}+{\frac{
u^2_-({\bm{k}})} {{\omega}+E({\bm{k}})-i{\Gamma}}}\right ].
\end{equation}
where $E={\sqrt{{\xi}^{2}_{K}+|{\Delta}|_{}^2}}$ and
$2u^2_{\pm}=1\pm {\xi}^{}_{K}/E$ are quasiparticle energy and
coherence factors, respectively. In accordance with (\ref{15}),
one should suppose that momentum-dependent quasiparticle weight
$z^{}_{\bm{k}}$ vanishes on the line of zeroes and corresponds to
a finite value $z$ $(0<z<1)$ on the PFC. Two terms can be referred
to pairs above and below the FC.

Diagonal Green function (\ref{19}) describes a non-superconducting
state with off-diagonal short-range (ODSRO) corresponding to the
existence of non-coherent ${\bm{K}}$~-~pairs above $T^{}_c$. Below
$T^{}_c$, the ODSRO transforms into the off-diagonal long-range
order (ODLRO) introduced by Yang. \cite{Yang}

Excitation with a transition from the bound paired state into
long-living QSS corresponds to quite small but finite decay
${\Gamma}={\Gamma}({\omega};{\bm{k}})$. The transitions into
stationary states above barrier energy $E^{}_b$ should be
associated with an infinitesimal decay, ${\gamma}\rightarrow +0$,
leading to conventional Fermi-liquid behavior of diagonal Gor'kov
function (\ref{19}) above $T^{\ast}_{str}$. Thus, a rise of QSS
results in a non-Fermi-liquid behavior of diagonal Green function
(\ref{15}) that can be manifested in rather wide temperature range
$T^{}_c <T \lesssim T^{\ast}_{str}$ relating to strong pseudogap
state. This range corresponds to transitions between boson-like
bound and quasi-stationary states. Therefore, Eq.~(\ref{19}) can
be considered as a bridge between the BCS and BEC approaches to
the problem of superconductivity, in accordance with the
assumption by Geshkenbein et al. \cite{GIL}

\section{Superconducting state with large momentum}

One can believe that the nearly nested FC of underdoped (up to
optimum doping) cuprate compound in the form of a square with
rounded corners, shown schematically in Fig.~3, results in the
fact that the ${\bm{K}}$~-~channel corresponding to an
incommensurate momentum ${\bm{K}}$ dominates the Cooper channel.
In such a case, it is a rise of coherence in the system of
${\bm{K}}$~-~pairs that determines SC transition temperature
$T^{}_{c}$, Therefore, there is a temperature range well below
$T^{}_{c}$ in which mean-field SC order parameter
${\Delta}^{}_c({\bm{k}})$, relating to the ${\bm{K}}$~-~channel,
can be approximately considered as governed by the only
self-consistency equation,
\begin{equation}\label{20}
{\Delta}^{}_c({\bm{k}})=-{\frac{1}{2}}\sum_{{\bm{k}}_{}^{\prime}}
{\frac{w({{\bm{k}},{\bm{k}}_{}^{\prime}}){\Delta}^{}_c({\bm{k}}_{}^{\prime})}
{E({\bm{k}}_{}^{\prime})}}
\tanh{\left({\frac{E({\bm{k}}_{}^{\prime})}{2T}}\right)}.
\end{equation}
Quasiparticle energy,
\begin{equation}\label{20'}
E({\bm{k}})={\sqrt{{\xi}_{K}^2({\bm{k}})+
|{\Delta}^{}_c({\bm{k}})|_{}^2+ |{\Delta}^{}_p({\bm{k}})|_{}^2}},
\end{equation}
aside from ${\Delta}^{}_c({\bm{k}})$, includes strong pseudogap
parameter, ${\Delta}^{}_p({\bm{k}})$, associated with QSS that can
exist above $T^{}_c$. Thus, Eq.~(\ref{20}) reflects the fact that
SC order arises from the state other than normal Fermi liquid.
Therefore, Eq.~(\ref{20}) differs from the conventional BCS
self-consistency equation. It is reasonable to assume that
${\Delta}^{}_p({\bm{k}})$ is small above optimum doping and
gradually increases with underdoping. This leads to smoothing of
the singularity in Eq.~(\ref{20}) and, as a result, to a gradual
decrease in $T^{}_c$ with underdoping.

One can reduce summation in Eq.~(\ref{20}) to small part
${\Xi}^{}_s$ of the domain of kinematic constraint similar to the
two-particle problem considered above. Under repulsive
interaction, a non-trivial solution to Eq.~(\ref{20}) can arise
due to a competition of positive and negative contributions to the
right-hand side of this equation. Thus, such a solution should
have a line of zeroes (nodal line, NL) inside ${\Xi}^{}_s$.

\begin{figure}
\includegraphics[]{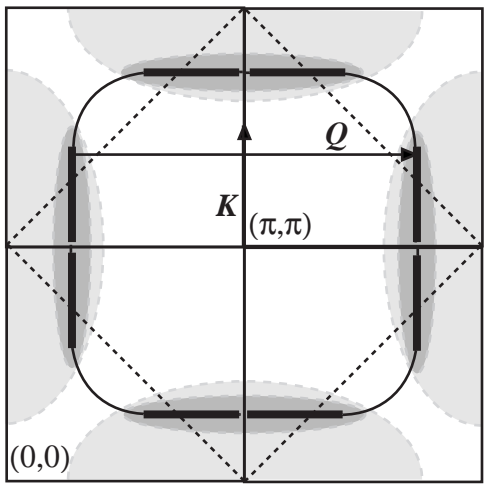}
\caption[*]{Schematic representation of the Fermi contour of
underdoped cuprate superconductor, in accordance with
Ref.\cite{Ino} Nested (with nesting momentum ${\bm{Q}}$) and
mirror nested (corresponding to total pair momentum ${\bm{K}}$)
pieces of the FC are shown by bold lines. Dotted line shows
magnetic Brillouin zone boundary. Shaded: dark narrow ovals
correspond to vicinities of Peierls singularity of screening;
light half-ovals designate extended saddle point
vicinities.}\label{FC.eps}
\end{figure}

In the case of phonon-mediated pairing with account of Coulomb
repulsion, \cite{BTSh} the NL of the order parameter coincides
with the boundary enclosing the domain of dynamic constraint. In
this domain, attraction dominates logarithmically weakened
repulsion in accordance with Eq.~(\ref{10}). This NL is disposed
everywhere outside of the FC, therefore, mirror nesting of the FC
can be considered as the only condition of the
${\bm{K}}$~-~pairing under attraction.

Peierls enhancement of the pairing interaction results in a strong
anisotropy of the NL disposed close to the FC inside ${\Xi}^{}_s$.
This corresponds to an increase in the dominant part of
${\Xi}^{}_s$ that mainly contributes into the logarithmic
singularity in the self-consistency equation and thus increases a
magnitude of ${\Delta}^{}_{c}$ (it is clear, that
$|{\Delta}^{}_{c}|$ should be much lesser than
${\varepsilon}^{}_0$).

Since the SC order parameter arising in the ${\bm{K}}$~-~channel
is essentially momentum-dependent, there are three characteristic
lines of zeroes: 1) the PFC on which kinetic energy of the pair
equals zero, $2{\xi}^{}_{K}({\bm{k}})=0$; 2) the NL of the order
parameter determined by ${\Delta}^{}_c({\bm{k}})=0$; 3) the curve
on which quasiparticle group velocity changes sign,
${\nabla}^{}_{\bm{k}}E({\bm{k}})=0$.

These three lines may have common points of intersection inside
${\Xi}^{}_s$, therefore, the NL can be disposed both above and
below the PFC. This results in qualitatively different
non-monotonic momentum dependence of coherence factors
$u^2_{\pm}({\bm{k}})$ for two kinds of directions in the
${\bm{k}}$~-~space intersecting at first the PFC and then the NL
and vice versa. Under pairing repulsion, the scattering across the
NL turns out to be dominating in comparison with scattering inside
or outside the NL in accordance with Suhl inequality Eq.(\ref{9}).

Due to the fact that $|{\Delta}^{}_{p}|_{}^2\neq 0$ in the strong
pseudogap state, coherence factors in diagonal Gor'kov function
(\ref{19}) may overlap each other near the PFC even above
$T^{}_c$. On the contrary, the BCS coherence factors are step-wise
functions without an overlap in the normal Fermi liquid state.

The SC state that arises below $T^{}_c$ should be described by
both diagonal and off-diagonal (anomalous) Gor'kov functions.
Taking into account the fact that mean-field (averaged over random
phases) pseudogap parameter ${\Delta}^{}_{p}$ vanishes whereas
mean-field SC condensate parameter ${\Delta}^{}_{c}\neq 0$ below
$T^{}_c$, one can introduce off-diagonal Gor'kov function
$F_{}^+({\omega};{\bm{k}})$ in a phenomenological way similar to
that we use to obtain diagonal Gor'kov function (\ref{19}). This
function describes the ODLRO state \cite{Yang} and can be written
as
\begin{equation}\label{22}
F_{}^+({\omega};{\bm{k}})=-{\frac{z^{}_{\bm{k}}
{\Delta}^{\ast}_{c}} {{({\omega}-E({\bm{k}})+i{\Gamma})}
{({\omega}+E({\bm{k}})-i{\Gamma})}}}.
\end{equation}
Factor $z^{}_{\bm{k}}$ is defined inside each of the crystal
equivalent domains of kinematic constraint ${\Xi}^{}_j$ where
$j=1,2,3,4$ in the case of tetragonal symmetry of cuprate planes.
Paired states with large total momenta ${\bm{K}}^{}_j$, both
coherent and non-coherent, arise exactly inside these domains.
Parameters ${\Delta}^{}_{cj}$, ${\Delta}^{}_{pj}$ and
${\Gamma}_{j}^{}$ are identical for any of ${\Xi}^{}_j$ differing
only by the domain of definition of ${\bm{k}}$.

In the whole of the Brillouin zone, the SC order parameter in the
mixed representation, ${\Delta}^{}_{cj}({\bm{R}},{\bm{k}})$, can
be presented as a superposition,
\begin{equation}\label{23}
{\Delta}^{}_{c}({\bm{R}},{\bm{k}})=\sum_{j=1}^4{\gamma}^{}_j({\bm{k}})
e_{}^{i{\bm{K}}^{}_j{\bm{R}}}{\Delta}^{}_{cj}({\bm{k}}),
\end{equation}
where ${\bm{R}}$ is center-of-mass radius-vector, coefficients
${\gamma}^{}_j({\bm{k}})$ should be chosen in accordance with the
symmetry of the order parameter.

As a function of ${\bm{R}}$, order parameter (\ref{23}) arising in
the ${\bm{K}}$~-~channel turns out to be spatially modulated
similar to Fulde-Ferrel-Larkin-Ovchinnikov (FFLO) state.
\cite{FF,LO} Such a modulation with relatively short wavelength
reflects a variation of ${\bm{K}}$~-~pair density and can be
associated with a pair density wave with a checkerboard order in
cuprate planes. Thus, the ${\bm{K}}$~-~pairing leads to a
microscopic ground of the pair density wave concept introduced
phenomenologically by Zhang. \cite{Zhang}

The SC order parameter can be expanded over the eigenfunctions of
pairing operator (\ref{7}), \cite{BK_UFN}
\begin{equation}\label{24}
{\Delta}^{}_{c}({\bm{R}},{\bm{k}})=\sum_s{\Delta}^{}_{cs}({\bm{R}})
{\varphi}^{}_s({\bm{k}}).
\end{equation}
In the simplest case, repulsion-induced SC pairing can be
described by two-component order parameter, so that two complex
components, ${\Delta}^{}_{cs}({\bm{R}})$, $s=1,2$, form the order
parameter structure in the framework of Ginzburg-Landau
phenomenology. Absolute values and relative phase of the
components are connected with the relative motion of
${\bm{K}}$~-~pair.

As follows from Ginzburg-Landau equation system for two-component
order parameter, \cite{BKSm} two qualitatively different SC states
become admissible. One of them corresponds to constant value $\pi$
of the relative phase of the components that is generic for
repulsion-induced superconductivity. The other state, with the
relative phase different from ${\pi}$, can be related to a change
of the phase of the wave function of ${\bm{K}}$~-~pair due to a
rise of internal magnetic field of spontaneous orbital currents.
Such currents can be associated \cite{BKSm} with insulating
orbital antiferromagnetic order, for example, in the form of DDW.
\cite{CLMN}

\section{Biordered superconducting state}

Superposition (\ref{23}) mixes, in particular, two paired states
with opposite momenta, ${\bm{K}}$ and $-{\bm{K}}$. It is clear
that particles composing pairs with these momenta can compose
pairs with zero momentum. Thus, the Cooper channel appears to be
associated with the ${\bm{K}}$~-~channel in accordance with
symmetry consideration.

In such a case, mean-field order parameters
${\Delta}^{}_0({\bm{k}})$ and ${\Delta}^{}_c({\bm{k}})$
corresponding to the Cooper and ${\bm{K}}$~-~channels,
respectively, should be the solution to a self-consistency
equation system. This system degenerates into two independent
equations (each of them determines one of the order parameters) if
one neglects the interconnection of the channels. Then, one would
obtain temperatures, $T^{}_c$ and $T^{\prime}_c$, of transitions
into the states with order parameters ${\Delta}^{}_c({\bm{k}})$
and ${\Delta}^{}_0({\bm{k}})$, respectively. Let us assume that
$T^{\prime}_c<T^{}_c$ even in the case when attractive
phonon-mediated pairing contributes to the Cooper channel. Then,
SC transition temperature $T^{}_c$ can be obtained directly from
Eq.~(\ref{20}).

At $T^{\prime}_c<T<T^{}_c$, there arises SC order due to
${\bm{K}}$~-~pairing with order parameter
${\Delta}^{}_c({\bm{k}})$ defined in relatively small vicinities
of the PFC where factor $z^{}_{\bm{k}}$ is close to unity. Inside
this temperature range, Cooper pairing on the whole of the FC
should be induced by ${\bm{K}}$~-~pairing. In this case, the
magnitude of ${\Delta}^{}_0({\bm{k}})$ has to be small in
comparison with the magnitude of ${\Delta}^{}_c({\bm{k}})$.

Thus, {\it{biordered}} SC state arising in such a way should be
described by two order parameters, ${\Delta}^{}_c$ and
${\Delta}^{}_0$, defined in the vicinities of the PFC and entire
FC, respectively. As temperature decreases from $T^{}_c$ down to
$T\approx T_c^{\prime}$, Cooper ordering with order parameter
${\Delta}^{}_0$ exists as induced by the ${\bm{K}}$~-~channel of
SC pairing. In this case, the superfluid density turns out to be
approximately proportional to the PFC length. Opening of the
Cooper channel at $T\approx T_c^{\prime}$ leads to a considerable
increase in ${\Delta}^{}_0$ and, as a result, in the superfluid
density which becomes proportional to the whole of the FC length
at $T\lesssim T_c^{\prime}$.

In the vicinities of the PFC, two branches ($m=1,2$) of strong
anisotropic quasiprticle spectrum of the biordered superconductor
can be written in the form
\begin{equation}\label{25}
E^{}_m({\bm{k}})={\sqrt{{\xi}^2_{K}({\bm{k}}) +|
{\Delta}^{}_p({\bm{k}})|_{}^2 +| {\Delta}^{}_c({\bm{k}}) \pm
{\Delta}^{}_0({\bm{k}})|_{}^2}}.
\end{equation}
Here, we take into account the fact that kinetic energy of
${\bm{K}}$~-~pair (\ref{4}) is equal to kinetic energy of Cooper
pair,
\begin{equation}\label{26}
2{\xi}^{}_0({\bm{k}})={\varepsilon}({\bm{K}}/2+{\bm{k}})+
{\varepsilon}(-{\bm{K}}/2-{\bm{k}})-2{\mu}.
\end{equation}
Two-gap spectrum (\ref{25}), with the lesser gap
$|{\Delta}^{}_c-{\Delta}^{}_0|$ observable at excitation energies
up to the greater gap $|{\Delta}^{}_c+{\Delta}^{}_0|$, should be
apparent at $T\lesssim T_c^{\prime}$. Above the greater gap the
spectral weight transfers from the low to high-energy branch of
quasiparticle spectrum.

Diagonal and off-diagonal Gor'kov functions of the biordered SC
state preserve their form, Eq.(\ref{19}) and (\ref{22}),
respectively, with the exception of the fact that SC order
parameter ${\Delta}^{}_c$ has to take into account both SC pairing
channels. Thus, these two channels result in two coexisting ODLRO
states.

Most likely, Cooper channel, including both Coulomb and
phonon-mediated pairing, cannot result in a rise of QSS.
Therefore, SC gap parameter turns out to be BCS-like everywhere on
the FC with the exception of the PFC on which the unconventional
${\bm{K}}$~-~channel is opened.

Symmetry of the biordered SC state is determined by Eqs.(\ref{23})
and (\ref{24}) where ${\Delta}^{}_s$ should be considered as the
components of momentum dependent order parameter arising as a
result of both Cooper and ${\bm{K}}$~-~pairing. One can
approximately represent the gap parameter in the conventional form
\begin{equation}\label{27}
{\Delta}({\bm{k}})=D({\bm{k}})(\cos{k^{}_x}\pm \cos{k^{}_y})
\end{equation}
where momentum dependent magnitude $D({\bm{k}})$ reflects mainly
strongly anisotropic contribution of the ${\bm{K}}$~-~pairing.
Upper (lower) sign in Eq.(\ref{27}) corresponds to extended
$s$($d$)~-~wave symmetry of the order parameter. It should be
noted that, in the case of Peierls enhanced ${\bm{K}}$~-~pairing,
nodal line of order parameter ${\Delta}({\bm{k}})$ may pass
through the center of the corresponding domain of kinematic
constraint. This results in the fact that order parameter
${\Delta}({\bm{k}})$ differs in sign on the opposite parts of the
PFC.

\begin{figure}
\includegraphics[]{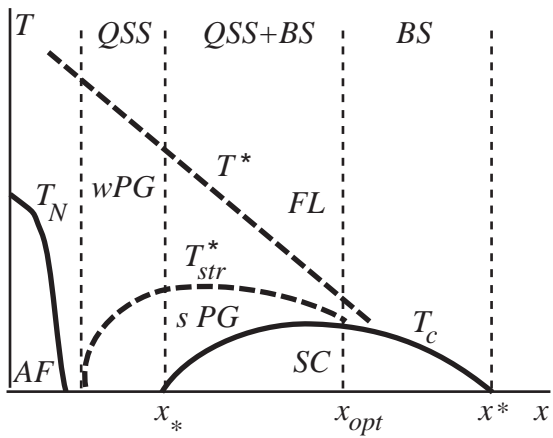}
\caption[*]{Schematic phase diagram of underdoped cuprates. Here,
$T^{}_{N}$ and $T^{}_c$ are phase transition temperatures
corresponding to Neel AF and unconventional SC orders,
respectively; $T^{\ast}_{str}$ is strong pseudogap (sPG) crossover
temperature, $T_{}^{\ast}$ bounds the region of weak pseudogap
(wPG) state. High doping and temperature state corresponds to
normal Fermi liquid (FL). On top, there are shown doping ranges in
which one can expect a rise of QSS, QSS together with BS and
BS.}\label{PhD.eps}
\end{figure}

\section{Conclusion}

We believe that biordered superconductivity may be generic for
such superconductors as doped cuprate compounds. Unconventional
features of these compounds, especially, universality of their
phase diagram (Fig.4), can be associated with evolution of the FC
and pairing interaction with doping.

It is clear that singular contribution to Coulomb pairing
interaction is sensitive to doping dependent form of the FC.
Therefore, one can suppose that oscillating real-space pairing
potential $w(r)$ varies with doping in such a way that only
noncoherent QSS of SC pairs arise under extremely low doping. This
corresponds to strong pseudogap penetrating into insulating region
of doping below the onset of superconductivity at $x=x^{}_{\ast}$.
In underdoped region $x^{}_{\ast}<x<x^{}_{opt}$, along with QSS,
there is a bound state. Both bound state energy $|E^{}_i|$ and QSS
decay ${\Gamma}$ increase with doping so that, near optimal doping
$x^{}_{opt}$, pair-break energy approximately coincides with the
energy corresponding to the loss of phase coherence. Thus, in
overdoped regime $x^{}_{opt}<x<x^{\ast}_{}$, pairing interaction
can result in only bound state.

As doping increases, the ${\bm{K}}$~-~channel may be dominated by
the Cooper channel and overdoped SC state can manifest properties
inherent in conventional BCS state. When doping exceeds
$x^{}_{opt}$, a decrease in $T^{}_c$ down to $T^{}_c=0$ at
$x=x^{\ast}_{}$ can be also associated with doping dependence of
the pairing interaction. This interaction becomes more repulsive
at $x>x^{}_{opt}$ as the FC gradually leaves the vicinity of the
extended van Hove saddle point.

A suppression of the phonon-mediated component of the SC pairing
may have the same origin, Effective increase in repulsion can
result in the fact that inequality (\ref{10}) can be reversed
because of not too large ratio $E^{}_F/{\varepsilon}^{}_D$ typical
of cuprates.

There is rather strong evidence that, in underdoped cuprates,
dimensionless ratio $2{\Delta}_{}^{(0)}/T^{}_c$ considerably
exceeds universal BCS value $3.52$. \cite{Oda} Here,
${\Delta}_{}^{(0)}$ has meaning of the SC energy gap extrapolated
down to $T=0$. In underdoped biordered superconductor, this
parameter should be determined by both Cooper and
${\bm{K}}$~-~pairing, therefore, one can assume that
${\Delta}_{}^{(0)}={\sqrt{{\Delta}^2_{} +{\Delta}^2_0
+{\Delta}^2_{p}}}$. Taking into account the fact that $T^{}_c$ is
determined by ${\bm{K}}$~-~pairing only and ${\Delta}^2_{}$,
${\Delta}^2_0$ and ${\Delta}^2_{p}$ are, generally speaking, of
the same order, one can easily conclude that ratio
$2{\Delta}_{}^{(0)}/T^{}_c$ may considerably exceed $3.52$
(observed in Ref.\cite{Oda} values
$2{\Delta}_{}^{(0)}/T^{}_c\gtrsim 10$). In overdoped regime,
strong pseudogap parameter ${\Delta}^{}_p \rightarrow 0$ and the
Cooper channel dominates ${\bm{K}}$~-~pairing. Therefore,
$2{\Delta}_{}^{(0)}/T^{}_c$ should be close to $3.52$ in
accordance with the BCS theory.

Superfluid density ${\rho}^{}_s$ should be determined by
condensation of ${\bm{K}}$~-~pairs within a broad temperature
range below $T^{}_c$, down to the onset of the Cooper channel.
Below $T_c^{\prime}$, superfluid density increases considerably;
conversely, off-condensate particle density decreases. Drude-like
behavior of the coherent contribution into optical conductivity
${\sigma}^{}_1(\omega)\sim {\omega}_{}^{-2}$, observed below
$T^{}_c$, can be connected with rather high off-condensate density
(experimental data available \cite{Basov} show that spectral
weight of the off-condensate particles may exceed spectral weight
of the SC condensate below $T^{}_c$). At $T\lesssim T_c^{\prime}$,
Drude component of the optical conductivity of biordered
superconductor, ${\sigma}^{}_1(\omega)$, should be suppressed due
to shedding of the off-condensate particles into the condensate of
Cooper pairs.

Two-gap excitation spectrum of biordered superconductor should be
consistent with tunnel conductance measurements \cite{Ved} and
quasi-linear temperature dependence of heat capacity,
$c^{}_V={\gamma}(T)T$. \cite{Loram}

All homologous cuprate series investigated demonstrate universal
dependence of $T^{}_c$ on the number of ${\text{CuO}}_2$ layers in
the unitary cell, $T^{}_c(n)$, with maximum at $n=3$. \cite{Scott}
Strong initial increase in $T^{}_c(n)$ cannot be associated with
local real-space pairing interaction. Weak interlayer tunnelling
can explain this feature qualitatively by rather small effective
enhancement of the coupling constant. \cite{ChKV} Coulomb pairing
with finite screening length ensures strong correlation between
electrons in the nearest-neighbor layers and results in a
quantitative explanation of $T^{}_c(n)$ leading to almost triple
increase of the coupling constant. \cite{BK_06}

Doping dependent isotope effect, observed in cuprate
superconductors, \cite {Khasanov} is highly sensitive to sample
quality and reflects the contribution of phonon-mediated component
to the SC pairing interaction. Depending on the interrelation
between Coulomb and phonon-mediated contributions, \cite{BKNT} the
exponent of the isotope effect on $T^{}_c$ can be close to both
zero, in the case of dominating repulsion, and BCS limit in the
opposite case of dominating phonon-mediated attraction. Relative
isotope shift, negligible above optimum doping, increases with
underdoping. We believe that this can be considered as an indirect
evidence in behalf of the fact that, in the case of low doping,
Coulomb correlation effects dominate phonon-mediated contribution
to the ${\bm{K}}$~-~channel which determines $T^{}_c$ in biordered
superconductor.

The isotope effect on the London penetration depth
${\lambda}^{}_L$, absent within the BCS theory, also turns out to
be enhanced with underdoping. \cite{Muller} The penetration length
is weakly sensitive to isotope substitution in a wide temperature
range well below $T^{}_c$. Then, starting from $T\approx
T_c^{\prime}$, the isotope shift on ${\lambda}^{}_L$ increases
gradually at $T\rightarrow 0$. As ${\lambda}^{-2}_L \sim
{\rho}^{}_s$, such a behavior of isotope effect on
${\lambda}^{}_L$ can be associated with temperature and doping
dependence od superfluid density inherent in the biordered SC
state.

\begin{acknowledgments}

We thank A.V. Chubukov, V.F. Elesin, V.L. Ginzburg, L.V. Keldysh,
and S.I. Vedeneev for very useful discussions. This work was
supported in part by the Russian Foundation for Basic Research
(project nos. 05-02-17077, 06-02-17186).

\end{acknowledgments}

\end{document}